\documentclass[traditabstract,letter]{myaa} 

\usepackage{graphicx}
\usepackage{txfonts}

\usepackage{epsf}
\usepackage{rotating}
\usepackage{graphics}

\def\ms{\hbox{\,m\,s$^{-1}$}}         %m.s -1
\def\m2s2{\hbox{\,m$^{2}$\,s$^{-2}$}} %m2.s -2
\def\kms{\hbox{\,km\,s$^{-1}$}}       %km.s -1
\def\vsini{\hbox{$v$\,sin\,$I$}}      %vsini
\def\Msun{\hbox{$\mathrm{M}_{\odot}$}}             %Msun
\def\Rsun{\hbox{$\mathrm{R}_{\odot}$}}
\def\Lsun{\hbox{$\mathrm{L}_{\odot}$}}
\def\Mjup{\hbox{$\mathrm{M}_{\rm Jup}$}}
\def\Rjup{\hbox{$\mathrm{R}_{\rm Jup}$}}

\def \1s{$1\,\sigma$}

\def \t0{T$_0$}

\def \cible{HD\,80606}
\def \sophie{{\it SOPHIE}}

\begin{document}
   \title{  Photometric and spectroscopic detection of the primary
transit of the 111-day-period planet \cible\,b\thanks{Based on observations made with 
the 1.20-m and 1.93-m telescopes at the Observatoire de Haute-Provence (CNRS), 
France, by the \sophie\ consortium (program 07A.PNP.CONS). 
Electronic tables of the data are available at the CDS via anonymous ftp to 
cdsarc.u-strasbg.fr (130.79.128.5) or via http://cdsweb.u-strasbg.fr/cgi-bin/qcat?J/A+A/}}

\author{
Moutou,~C. \inst{1}
\and H\'ebrard,~G. \inst{2}
\and Bouchy,~F. \inst{2,3}
\and Eggenberger,~A. \inst{4}  
\and Boisse,~I. \inst{2}
\and Bonfils,~X. \inst{4}
\and Gravallon,~D. \inst{3}
\and Ehrenreich,~D. \inst{4}
\and Forveille,~T. \inst{4} 
\and Delfosse,~X. \inst{4}
\and Desort,~M. \inst{4}
\and Lagrange,~A.-M. \inst{4}
\and Lovis,~C. \inst{5}
\and Mayor,~M. \inst{5}
\and Pepe,~F. \inst{5}
\and Perrier,~C. \inst{4} 
\and Pont,~F. \inst{6}
\and Queloz,~D. \inst{5}
\and Santos,~N.C. \inst{7}
\and S\'egransan,~D. \inst{5}
\and Udry,~S. \inst{5}
\and Vidal-Madjar,~A. \inst{2}
}

\institute{
Laboratoire d'Astrophysique de Marseille, UMR 6110, CNRS\&Univ. de Provence, 38 rue Fr\'ed\'eric Joliot-Curie, 13388 Marseille cedex 13, France
\email{Claire.Moutou@oamp.fr}
\and Institut d'Astrophysique de Paris, UMR7095 CNRS, Universit\'e Pierre \& Marie Curie, 98bis boulevard Arago, 75014 Paris, France
\and Observatoire de Haute-Provence, 04870 Saint-Michel l'Observatoire, France
\and Laboratoire d'Astrophysique, Observatoire de Grenoble, Universit\'e J. Fourier, BP 53, 38041 Grenoble, Cedex 9, France
\and Observatoire de Gen\`eve, Universit\'e de Gen\`eve, 51 Chemin des Maillettes, 1290 Sauverny, Switzerland
\and School of Physics, University of Exeter, Exeter, EX4 4QL, UK 
\and Centro de Astrof\'isica, Universidade do Porto, Rua das Estrelas, 4150-762 Porto, Portugal
}
   \date{Received ; accepted }
 
  \abstract{  
We report the detection of the primary transit of the extra-solar planet 
\cible\,b, thanks to photometric and spectroscopic observations performed 
at the Observatoire de Haute-Provence, simultaneously with the CCD 
camera at the 120-cm telescope and the \sophie\ spectrograph on the 193-cm 
telescope.  We observed the whole egress of the transit and 
partially its central part, in both datasets with the same timings. The ingress occurred before 
sunset so was not observed. The full duration of the transit was between 9.5 
and 17.2~hours. The data allows the planetary radius to be measured
($R_{\mathrm{p}} = 0.9 \pm 0.10\,{\rm R}_{\rm Jup}$) and other 
parameters of the system to be refined. Radial velocity measurements show the 
detection of a prograde Rossiter-McLaughlin effect, and provide a hint of a spin-orbit 
misalignment. 
If confirmed, this misalignment would corroborate the hypothesis that 
\cible\,b owes its unusual orbital configuration to Kozai migration.
\cible\,b is by far the transiting planet on the longest period 
detected today. Its unusually small radius reinforces the observed relationship 
between the planet radius and the incident flux received from the star and opens new questions for theory.
Orbiting a bright star ($V=9$), it opens opportunities for 
numerous follow-up studies.
}


\titlerunning{Primary transit of \cible\,b}
\authorrunning{C. Moutou et al.}
 
\maketitle

%
%________________________________________________________________

\section{Introduction}

The extra-solar planet \cible\,b was discovered with the ELODIE spectrograph by 
Naef et al.~(\cite{naef01}). This is a giant planet (4\,M$_\mathrm{Jup}$)
with a 111-day orbital period on an extremely eccentric orbit ($e=0.93$). 
\cible\ is a member of a binary system 
(with HD\,80607) with a separation
of $\sim1200$\,AU ($\sim20^{\prime\prime}$ on the sky). Wu \& 
Murray~(2003) suggest that the present orbit of \cible\,b results from the
combination of the Kozai cycles (induced by the distant stellar
companion) and tidal friction.
Recently, Laughlin et al.~(\cite{laughlin09}) report detecting 
a secondary transit for \cible\,b using 8 $\mu$m Spitzer 
observations around the periastron passage. This implies an inclination 
of the system near $i=90^{\circ}$, and a $\sim15$\,\% probability that 
the planet also shows primary transits. 

If transits occur, opportunities for detecting them are rare because the orbital 
period of the planet is almost four months.  We managed an observational 
campaign to attempt detection of the transit of \cible\,b
scheduled to happen on Valentine's Day (14 February 2009). We 
simultaneously used instruments of two telescopes of the Observatoire de 
Haute-Provence (OHP), France: the CCD camera at the 120-cm telescope, and the 
\sophie\ spectrograph at the 1.93-m telescope. This allows us to report the 
detection of the primary transit of \cible\,b, in photometry as well as in 
spectroscopy through the Rossiter-McLaughlin effect. 

We recall the stellar characteristics of the primary star, which we used in our analysis of the transit data presented hereafter: 
\cible\ is a G5-type star with a parallax measured by Hipparcos of 17$\pm$5 mas. A compilation of spectroscopic data from the literature gives an effective temperature of 
$5574\pm50$~K, $\log g$ of $4.45 \pm0.05$ and a high metallicity of 0.33~dex (Santos et al. \cite{santos04}, Valenti \&Fischer \cite{valenti}). The stellar mass can be 
estimated using isochrones and we get $0.98 \pm 0.10 \,\Msun$. Using the relationship between luminosity, temperature, gravity, and mass, the stellar luminosity is estimated 0.84 $\pm$ 0.13 \Lsun. Finally, relating radius with luminosity and temperature, we derive a radius of $0.98 \pm 0.07 \, \Rsun$. These mean values were obtained after several iterations over mass and radius determination.
Chromospheric activity gives an age interval of 1.7 to 7.6 Gyrs (Saffe et al.~\cite{saffe}).

\section{The photometric transit of HD\,80606b}
\begin{figure}[h] 
\begin{center}
\includegraphics[width=8.5cm,angle=0]{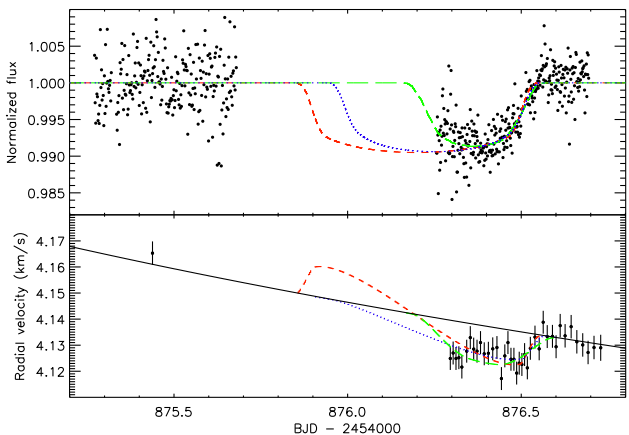}
\caption{Photometry (top) and radial velocities (bottom) of \cible\ from 12 to 14 
February, 2009, obtained at OHP with the 120-cm and 193-cm telescopes, 
respectively.  The planetary transit is detected in both datasets at the same timing. \textit{Top:} Superimposed 
are the two extreme ($b=0$ in red-dashed, and $b=0.91$ in green-long-dashed) and the mean 
(in blue-dotted, $b=0.75$) models that correspond to our data set. 
\textit{Bottom:} The orbital solution is overplotted (solid line, Table~\ref{table_parameters}), 
together with Rossiter-McLaughlin effect models presented in Table~\ref{tabross}, 
in red-dashed ($b=0$, $\lambda=0^{\circ}$), 
blue-dotted ($b=0.75$, $\lambda=63^{\circ}$), and
green-long-dashed lines ($b=0.91$, $\lambda=80^{\circ}$).
}
\label{photom1}
\end{center}
\end{figure}

Predicted epochs for the primary transit of HD\,80606b were given by  Laughlin et al.~(\cite{laughlin09}). We carried observations around the expected transit epoch JD=2,454,876.5 with the 120-cm telescope at OHP, equipped with a 12~arcmin $\times$ 12~arcmin CCD camera. The Bessel $R$ filter and a neutral density were inserted, to insure unsaturated, focused images of the $V=9$ target. We obtained 326 frames on 13 Feburary 2009 and 238 frames on the preceeding night, for comparison. Typical exposure times range between 60 sec on the first night and 20-30 sec on the second. Aperture photometry was then performed on both data sequences. Apertures of 8 and 6 pixels were used for the first and second observing nights, respectively. The secondary companion HD\,80607 is taken as a reference for HD\,80606. Both stars are separated by 24 pixels, which prevents contamination even using simple aperture photometry. The sky background is evaluated in rings of about 12-15 pixel radii. The resulting lightcurve is shown in Fig.~\ref{photom1} (upper panel) with all data included. The data quality is significantly better during the transit night, because of different seeing conditions. The $rms$ is about 0.0023 and 0.0030, respectively. 

An egress is clearly detected in the data sequence obtained during the night 13-14 February. A shift of almost one half transit is observed, in comparison to expected ephemeris. 
Long-term systematics are observed in the lightcurve and removed by a polynomial function of the airmass, with the criterion of getting a flat section of the out-of-transit flux. This correction does not strongly affect the transit shape. It is checked on the 12-13 February sequence that long-term fluctuations are low (not corrected for in Fig. \ref{photom1}). The beginning of the transit sequence unambiguously shows that we do not detect the ingress of the transit. The first hour of the sequence indicates a slight decrease but the data are quite noisy due to the low object's elevation, and this may be introduced by the correction for airmass variations. We observed in total 7 hours during transit on the second night, and 3.4 hours after the transit. In addition, we gathered 9.8 hours out of transit on the first night.

\begin{figure}[h] 
\begin{center}
\includegraphics[height=8.5cm,angle=90]{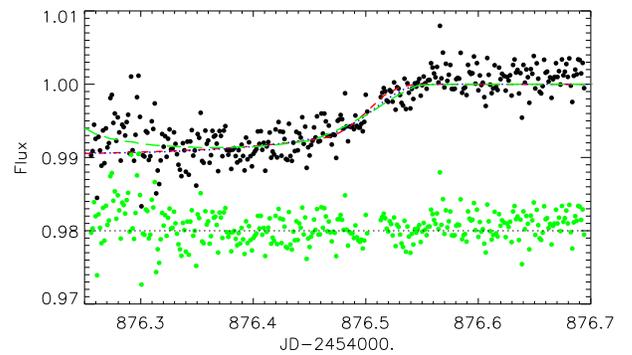}
\caption{Zoom-in plot on the photometric transit and the three models with the impact parameter ranging from 0 to 0.91. The residuals are shown below, with an offset of 0.02 on the Y-axis for clarity, and they correspond to the mean model with $b=0.75$.}
\label{fig1}
\end{center}
\end{figure}

Modelling the primary transit lightcurve of \cible\,b is done in the first place using models of circular orbits, to constrain the inclination and the radius ratio. The Universal Transit Modeler (Deeg \cite{deeg}) is used, including the limb-darkening coefficients of Claret (2000) for the $r'$ filter, and parameters of the orbits given in the next section. Figures ~\ref{photom1} and \ref{fig1} show three transit models superimposed to the data, corresponding to impact parameter $b$ ranging from 0 to 0.91 or inclinations ranging from 89.2 to 90.0 deg. The $O-C$ residuals depicted in Fig.Ê\ref{fig1} correspond to an average transit duration of 13.5 hours, with $b=0.75$ and $rms$ of 0.0023. The transit depth imposes a radius ratio $R_\mathrm{p}/R_*$ of $0.094 \pm 0.009$. The planet radius is then estimated  to be $0.9\pm0.10\,\Rjup$. To match the full transit lightcurve, a model including eccentricity would be required. The asymmetry of the ingress and egress should be detected and properly fitted, for instance. Since we have a partial transit, the approximation of the circular modelling is acceptable here, if one takes the relative projection of the transit angle and the line of sight into account. In a further study, we plan to investigate the modelling of the asymetric transit by including the eccentric orbit. We do not expect major differences compared to the simple fit performed here, before new, more complete photometric data are obtained.

\section{The spectroscopic transit of HD\,80606b}

We observed \cible\ with the \sophie\ instrument at the 1.93-m telescope of OHP. 
\sophie\ is a cross-dispersed, environmentally stabilized echelle spectrograph 
dedicated to high-precision radial velocity measurements  (Bouchy et al.~\cite{bouchy06}; 
Perruchot et al.~\cite{perruchot08}).
We used the high-resolution mode (resolution power $R=75,000$) of the 
spectrograph and the fast-readout mode of the CCD detector.
The two  3''-wide circular apertures (optical fibers) were used, the first one  
centred on the target and the second one on the sky to simultaneously 
measure its background.  This second aperture, 2' away from the first one, 
allows us to check that there is no significant pollution due to moonlight
on the spectra of the target.
We obtained 48 radial velocity measurements from 8 to 17 
February 2009, including a full sequence during the night 13  
February (BJD$\;=54876$), when the possible transit was expected to occur 
according to the ephemeris. The exposure times range from 600 to 1500 sec, insuring a 
constant signal-to-noise ratio. This observation was performed in parallel to the photometric ones.

The sequence of the transit night is plotted in Fig.~\ref{photom1}, lower panel, together 
with the measurement secured the previous night. The Keplerian curve expected 
from the orbital parameters is overplotted. The radial velocities of the 13 February 
night are clearly blue-shifted by $\sim10$\,\ms\ from the Keplerian curve in the 
1st half of the night, then match the Keplerian curve in the 2nd half of the night.
This is the feature expected for a planet transiting on a prograde orbit, 
according to the Rossiter-McLaughlin (RM) effect.
This effect occurs when an object transits in front of a rotating star, causing a spectral 
distortion of the stellar lines profile, and thus resulting in a Doppler-shift 
anomaly (Ohta et al. \cite{otha}; Gim\'enez et al. \cite{gimenez06b}; Gaudi \& 
Winn \cite{gaudi}). 

On the RM feature of \cible\,b (Fig.~\ref{photom1}), 
the third and fourth contacts occurred at BJD$\;\simeq54876.45$ and 
BJD$\;\simeq54876.55$, respectively,
whereas the 
first contact occurred before sunset and was not observed. These timings agree
with those of the photometry (Sect. 2 and Fig.~\ref{fig1}) and
the detection of the Rossiter-McLaughlin anomaly is unambiguous. 

The Keplerian curve in Fig.~\ref{photom1} corresponds to the orbital parameters that 
we refined for \cible\,b. We used the \sophie\ measurements performed out of the 
transit, as well as 45 Keck measurements (Butler et al. 2006) and 
74 ELODIE measurements (55 published by Naef et al.~(\cite{naef01}) 
and 19 additional measurements obtained from BJD$\;=52075$ to 52961). 
We allowed free radial-velocity shifts between the three datasets.
We used the constraint of the secondary transit given by Laughlin et 
al. (\cite{laughlin09}) ($T_e=2\,454\,424.736\pm0.003$~HJD). 
We also used our constraint on the primary transit considering 
that the end of transit is $T_{egress} = 2\,454\,876.55\pm0.03$~BJD. 
%and that the minimal duration is 12 hours. 
From these contraints, 
we estimated that the inclination of the system is from 90$^{\circ}$  
($T_t=2\,454\,876.20$ BJD with 17.2 h duration) to 89.2$^{\circ}$ 
($T_t=2\,454\,876.32$ BJD with 9.4 h duration).

\begin{figure}[t] 
\begin{center}
\vspace{1cm}
\includegraphics[scale=0.45,angle=-0]{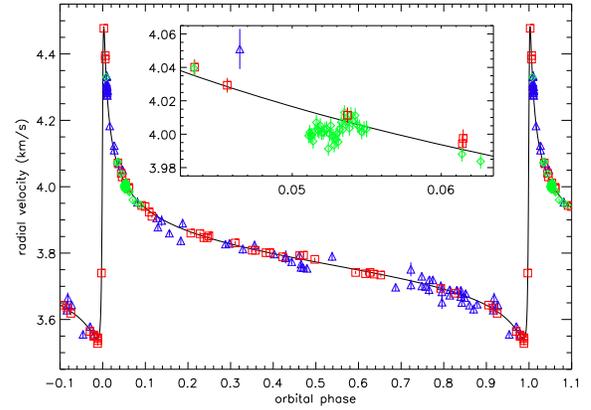}
\caption{Phase-folded radial velocity measurements of \cible\ 
as a function of the 
orbital phase, and Keplerian fit to the data. ELODIE 
data in blue, Keck data in red, \sophie\ data in green.
Orbital parameters corresponding to this 
fit are reported in Table~\ref{table_parameters}. The inset shows a zoom around the transit phase. One Keck spectrum was obtained 1hr after transit.}
\label{fig_orb_phas}
\end{center}
\end{figure}

Assuming those constraints, 
we adjusted the Keplerian orbit.   
The dispersion of the radial velocities around this fit is 8.6~\ms, and 
the reduced $\chi^2$ is 1.4.
The obtained parameters are reported in Table~\ref{table_parameters}. They 
agree with those of Laughlin et al.~(\cite{laughlin09}), except for the period, 
where there is a 3-$\sigma$ disagreement. The full data set and orbital solution are plotted in 
Fig.~\ref{fig_orb_phas}. We note that no anterior data was obtained during the transit by any instrument, as shown in the inset of Fig.~\ref{fig_orb_phas}.

\begin{table}[b]
  \caption{Fitted orbit and planetary parameters for \cible\,b.}
  \label{table_parameters}
\begin{tabular}{lcc}
\hline
\hline
Parameters & Values and 1-$\sigma$ error bars & Unit \\
\hline
$V_r$  (Elodie)                         & $3.788\pm 0.002$                      &   \kms        \\
$V_r$  (\sophie)                         & $3.911\pm 0.002$                      &   \kms        \\
%$V_r$   (Keck)                        & --                    &   --        \\
$P$                                     & $111.436\pm0.003$                         &   days        \\
$e$                                     & $0.934\pm0.003$                               \\
$\omega$                        & $300.6\pm0.4$                                   &   $^{\circ}$ \\
$K$                                     & $472\pm5$                                     &   \ms \\
$T_0$ (periastron)              & $2\,454\,424.857\pm0.05$                    &   BJD         \\
$M_\mathrm{p} \sin i$           & $4.0 \pm 0.3$$^\dagger$              &   M$_\mathrm{Jup}$ \\
$a$                                     & $0.453 \pm 0.015$$^\dagger$             &  AU \\
$T_t$ (primary transit)              & $2\,454\,876.27\pm0.08$                    &   BJD         \\
$T_e$ (secondary transit)              & $2\,454\,424.736\pm0.003$$^\ddag$                  &   BJD         \\
%transit dur. ($1^\mathrm{st}$ to $4^\mathrm{rd}$ cont.) &  & h \\
%f. transit dur. ($2^\mathrm{nd}$ to $3^\mathrm{rd}$ cont.) &  & h \\
t$_{14}$		&	$9.5-17.2$ & hours\\
t$_{23}$		& 	$8.7-15.7$ & hours\\
$M_\star$				&	$0.98\,\pm0.10$  				&   M$_\odot$	 \\
$R_\star$				&	 $0.98\,\pm0.07$			&   R$_\odot$ \\
$R_\mathrm{p}/R_*$ & $0.094 \pm 0.009$ \\
$R_\mathrm{p}$			&	$0.9\,\pm0.10$ 		&   R$_\mathrm{Jup}$ \\
$b$			&	0.75 (-0.75, +0.16) & \\
$i$                            &        $89.6 (-0.4,+0.4)$ &   $^{\circ}$\\
%$\lambda$			&	 		&   $^{\circ}$\\
\hline
$\dagger$: using $M_\star = 0.98\,\pm0.10$\,M$_\odot$\\
$\ddag$: from Laughlin et al.~(\cite{laughlin09})
\end{tabular}
\end{table}

To model the RM effect, we used the analytical approach 
developed by Ohta et al. (\cite{otha}). The complete model has 12 parameters: 
the six standard orbital parameters, the radius ratio $r_\mathrm{p}/R_*$, the orbital semi-major axis 
to stellar radius $a/R_*$ (constrained by the transit duration), the sky-projected angle between 
the stellar spin axis and the planetary orbital axis $\lambda$, the sky-projected stellar rotational 
velocity \vsini, the orbital inclination $i$, and the stellar limb-darkening coefficient $\epsilon$. 
For our purpose, we used the orbital parameters and photometric transit parameters as 
derived previously. 
We fixed the linear limb-darkening coefficient $\epsilon=0.78$, based on Claret (\cite{claret}) 
tables for filter $g'$ and for the stellar parameters derived in Sect.~1.
Our free parameters are then $\lambda$, {\vsini}, and $i$. 
As we observed a partial transit, there is no way to put a strong constraint 
on the inclination $i$. 
We then decided to adjust $\lambda$ for different values of $i$ in the range 
89.2 - 90 $^{\circ}$. 
The results of our fits (Table~\ref{tabross} and Fig.~\ref{photom1}, lower panel) first 
show that the stellar rotation is prograde relative to the planet orbit. 
Assuming $i=90^{\circ}$ and $\lambda=0^{\circ}$, 
the projected rotation velocity of the star 
\vsini\  determined by our RM fit is 2.2~{\kms}. This agrees with the value 1.8\,\kms\ obtained 
by Valenti \& Fisher~(2005), as well as our spectroscopic determination (2-3\,\kms) from SOPHIE 
spectra. This latest one could be slightly overestimated here due to the high metallicity of \cible.
We decided to fix this value and to explore the different values of inclination angle $i$ to estimate 
the spin-orbit $\lambda$ angle.  
We see in Table~\ref{tabross} that, if the transit is not central, then the RM fit 
suggests that the spin-orbit angle is not aligned. 

\begin{center}                        
\begin{table}[h]
\caption{Parameter sets for the Rossiter-McLaughlin effect models.}            
\label{tabross}     
\begin{tabular}{l c c c c }      
\hline
\hline                
  $i$   & transit duration  & Ttransit  &  spin-orbit $\lambda$ & $\chi^2$  \\  
  (deg) & (hours)           &   BJD -2454000       & (deg)                &        \\  
\hline                       
90.0   &  17.2 &  876.20  & 0   & 33.0 \\
89.6   &  15.5 &  876.24  & 63  & 25.3 \\
89.2   &   9.4 &  876.32  & 86  & 33.4 \\
\hline                                   
\end{tabular}
\end{table}
\end{center}

\section{Discussion and conclusion}

Despite the low probability of a 111-day period system being seen edge-on at both at the 
primary and the secondary transit phases (about 1\,\% in the case of \cible\,b), the data 
acquired at the Observatoire de Haute-Provence and presented here show 
this alignment unambiguously. With a partial transit observed, we were able to constrain the orbital parameters, including the inclination 
with a precision of $\simeq0.4^{\circ}$, and to measure the planetary radius. 
The error bars of our measurements should be taken with caution, however, and the system's parameters slightly revised when more complete data is obtained, since systematic noise is more difficult to correct with incomplete transits. 

\begin{figure}[h]
\begin{center}
\includegraphics[width=8.5cm,angle=0]{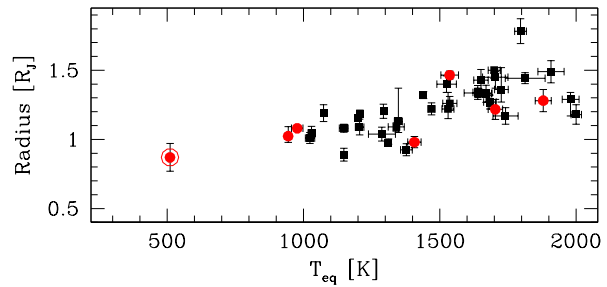}
%\caption{Mass-radius diagram for transiting exoplanets. Planets with orbital 
%periods longer than five days are plotted in red (HAT-P-2b, WASP-8b, CoRoT-Exo-6b, %CoRoT-Exo-4b, 
%HD\,17156b and \cible\,b.)}
\caption{Radius of transiting gas giant planets as a function of the equilibrium temperature  ($T_{eq} \sim T_* (R_*/a)^{1/2} (1-e^2)^{-1/8}$). HD80606 is circled.  Its position reinforces the correlation between incident flux and radius. All transiting gas giants are included above 0.4 \Mjup. The red circles show the planets with mass higher than 2 \Mjup. They follow the same tendency as Jupiter-like planets (black squares).}
\label{mr2009}
\end{center}
\end{figure}

The planet has a low radius ($0.9\,\Rjup$) considering its mass ($4\,\Mjup$). 
Since it is also by far the transiting gas giant receiving the lower irradiation from its parent star, it is tempting to see its small radius as reinforcing the explanation of anomalously large hot Jupiters as caused primarily to stellar irradiation, as proposed for instance by Guillot \& Showman (2002). Figure \ref{mr2009} shows the increasingly clear correlation between equilibrium temperature and size for transiting gas giants. The relation holds for both massive planets ($>$ 2 \Mjup) and Jupiter-like planets. 
Tidal effects (Jackson et al., 2007) may play a role in the observed radius of \cible\,b. 
The high metallicity of the parent star also helps provide refractory material for a massive core, although the required enrichment would be beyond actual expectations. More theoretical development is needed to reproduce the system's parameters, taking the whole history of orbital evolution and variations in the irradiation conditions into account. 

Most of the $\sim60$ known transiting planets are orbiting close to their hosting stars. 
Only 5 of them have periods longer than 5 days, the most distant from its star being HD\,17156b, 
on a 21.2-day period. \cible\,b has
% is the sixth detected transiting planet above 5-day period, with 
by far the longest period (111.4 days). 
It may be compared to other planetary systems with a massive planet in an eccentric orbit: 
HD\,17156, HAT-P-2, and XO-3. \cible\,b has a smaller radius than those planets, which can be 
related to the migration history or to the changes in stellar irradiation along the orbit 
(Laughlin et al, 2009). The shape of the Rossiter-McLaughlin anomaly shows that the orbit 
of \cible\,b is prograde, and suggests that it could be significantly 
inclined relative to the stellar equator. 
%A spin-orbit misalignment is expected if HD\,80606\,b owes its current orbital
%configuration to Kozai migration (Wu \& Murray 2003). Kozai migration 
%can explain the formation of the planet only if the initial relative inclination 
%of the system is large. % ($\gtrsim$85^{\circ}$).
Since a high initial relative inclination is a key requirement for Kozai migration to work (Wu \& Murray 2003), this observation is not surprising. Tighter constraints on the spin-orbit misalignment in \cible\, may support the Wu \& Murray formation scenario and may provide compelling evidence that the orbital evolution of \cible\,b was once dominated by the binary companion.
Among the 11 other
transiting planets with Rossiter-McLaughlin measurements, the only
system to show a significant spin-orbit misalignment is XO-3,
another massive and eccentric planet (H\'ebrard et al.~\cite{hebrard08}; Winn 
et al.~\cite{winn09}). HAT-P-2b is aligned (Winn et al.~\cite{winn07}; 
Loeillet et al. \cite{loeillet07}).

\cible\,b is thus a new Rosetta stone in the field of planetary transits.
By orbiting a bright star ($V=9$), it opens opportunities for numerous follow-up studies, including: observation of a full photometric transit from space or multi-site campaigns to measure a complete spectroscopic transit sequence.

After submission, we learned that other observations of this system had been obtained on the same day. Fossey et al (2009) and Garcia-Melendo \& McCullough (2009) independently confirm the detection of the photometric transit. In addition, MEarth observations (D. Charbonneau, priv. comm. and oklo.org) held in Arizona show a flat lightcurve of \cible, which limits the transit duration to less than 12 hours, hence reinforcing evidence of a spin-orbit misalignment (solutions of red and blue models in Fig.~\ref{photom1} are rejected). The grazing eclipse configuration would also result in a slightly larger planetary radius. Further analyses will follow in a forthcoming paper.

\begin{acknowledgements}
We are extremely grateful to Greg Laughlin for calling attention to the potential transit and
encouraging observations.
We thank the technical team at Haute-Provence Observatory for their support 
with the \sophie\ instrument and the 1.93-m and 1.20-m telescopes.
Financial support for the \sophie\ Consortium 
from the "Programme national de plan\'etologie" (PNP) of CNRS/INSU, 
France, and from the Swiss National Science Foundation (FNSRS) 
are gratefully acknowledged. 
We also acknowledge support from the French National Research Agency~(ANR). 
N.C.S. would like to thank Funda\c{c}\~ao 
para a Ci\^encia e a Tecnologia, Portugal, for the support through programme Ci\^encia 2007 and project grant 
reference PTDC/CTE-AST/66643/2006.
\end{acknowledgements}

\end{document}